\documentclass[letterpaper]{article}
\usepackage{spconf,amsmath,amssymb,graphicx,xcolor}

\newcommand{\mypar}[1]{{\bf #1.}}

\title{Accelerating Sparse Ternary GEMM for Quantized ML on Apple Silicon}
\name{%
  Muhammed Bilal*\thanks{*Equal contribution.}\quad
  Baraq Lipshitz*\quad
  Charalampos Maraziaris*\quad
  Alessio Melone*%
}

\address{Department of Computer Science\\ ETH Zurich, Switzerland}

\begin{document}
%
\maketitle
%


\begin{abstract}
Sparse Ternary General Matrix-Matrix Multiplication (GEMM) remains under-optimized in existing libraries for Apple Silicon CPUs. We present a Sparse Ternary GEMM kernel optimized specifically for Apple’s M-series processors. We propose a set of architecture-aware optimizations, including a novel blocked and interleaved sparse data format to improve memory locality, strategies to increase Instruction-Level Parallelism (ILP), and NEON-based Single Instruction Multiple Data (SIMD) vectorization to exploit data-level parallelism. Our scalar implementation achieves up to a 5.98× performance increase over a traditional Ternary Compressed Sparse Column (TCSC) baseline for large matrices with 50\% ternary nonzero values (sparsity), reaching up to a 50.2\% of the processor's theoretical peak performance, and remains stable across varying sparsity levels. Our vectorized implementation delivers up to a 5.59× performance increase for large matrices with 25\% sparsity, and remains stable across varying sparsity levels.

\end{abstract}

\section{Introduction}\label{sec:intro}

\mypar{Motivation} 
The compute needed by Large Language Models (LLMs) presents a significant barrier to both their training and efficient deployment in cloud and consumer hardware. To address this, model compression techniques such as quantization and pruning have become essential. A particularly effective approach combines these methods by quantizing model weights to a ternary set of values \{-1, 0, +1\}, which simultaneously introduces high levels of sparsity. This transforms the fundamental neural network operation from General Matrix-Matrix Multiplication (GEMM) into Sparse Ternary GEMM. The performance of this kernel is critical for low-latency inference. However, standard linear algebra libraries are not optimized for this specific task on Apple Silicon. In this project, we create a highly optimized implementation of Sparse Ternary GEMM tailored for ARM-based Apple Silicon architecture. This kernel is useful across Apple's entire product line, specifically tuned for M-series processors.

\mypar{Contribution}
We present a scalar and an associated vector implementation for fast Sparse Ternary GEMM tailored to Apple's M-series processors. Our best scalar algorithm performs up to 5.98x faster than the naive TCSC baseline, achieving consistently 50\% peak performance on Apple M1.

\section{Background on the Algorithm}\label{sec:background}
We focused on optimizing:
\begin{equation}
\underbrace{Y}_{M \times N} = \underbrace{X}_{M \times K} \overbrace{W}^{\in \{1,0,-1\}^{K \times N}} + \underbrace{b}_{N}
\end{equation}
where $X$ is dense and $b$ is broadcast-added to each row of the product.

To efficiently perform this matrix multiplication, the sparse matrix \(W\) is stored in a variant of the Compressed Sparse Column (CSC) format, called Ternary Compressed Sparse Column (TCSC). This format is specifically designed for ternary matrices, where zero values are omitted. Unlike traditional CSC which stores both values and indices, TCSC exploits the ternary nature by separating the non-zero elements of +1 and -1 into distinct data structures.

The TCSC format consists of four integer arrays:
\begin{itemize}
    \item \texttt{col\_start\_pos}: column start pointers for +1s
    \item \texttt{col\_start\_neg}: column start pointers for -1s  
    \item \texttt{row\_index\_pos}: row idx of all +1s, column-wise
    \item \texttt{row\_index\_neg}: row idx of all -1s, column-wise
\end{itemize}

\begin{sloppypar}
For each column $j$, positive values are located at 
\texttt{row\_index\_pos[col\_start\_pos[j]:}\allowbreak
\texttt{col\_start\_pos[j+1]]}, while negative values are found at 
\texttt{row\_index\_neg[col\_start\_neg[j]:}\allowbreak
\texttt{col\_start\_neg[j+1]]}.
This organization eliminates the need to store explicit values since the sign is implicitly encoded by the array choice, reducing memory overhead compared to traditional sparse formats.
\end{sloppypar}

\begin{figure}[htbp]
	\centering
	\includegraphics[width=0.5\textwidth]{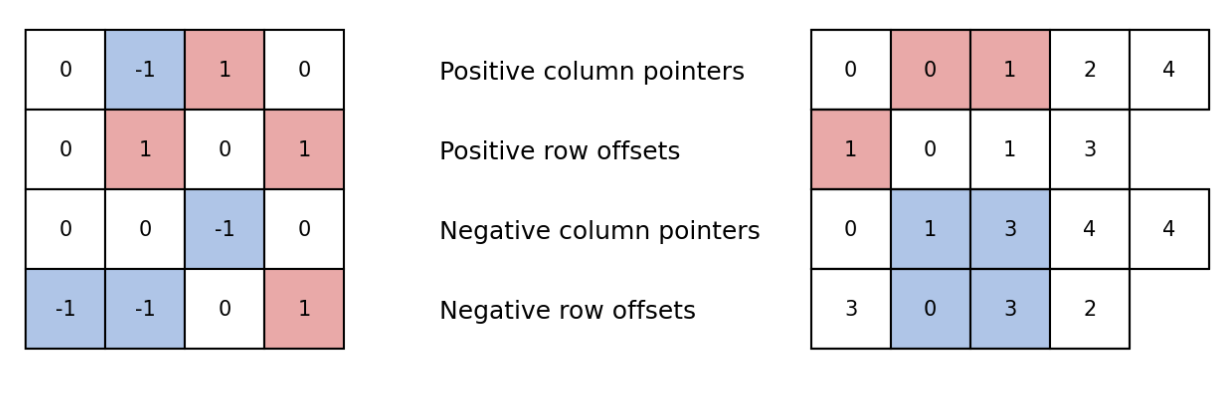}
	\caption{Example of TCSC format}
	\label{fig:base_tcsc}
\end{figure}
The BaseTCSC algorithm implements the core matrix multiplication using the TCSC format. It iterates through each output element $Y[m][n]$ and computes the dot product between row $m$ of matrix $X$ and column $n$ of the sparse matrix $W$. For each column $n$, the algorithm processes positive values first (adding $X[m][row\_index\_pos[k]]$ to the accumulator), followed by negative values (subtracting $X[m]$ $[row\_index\_neg[k]]$).

This approach has two key characteristics: it maintains good spatial locality for matrix $Y$ (accessed sequentially, element by element), and also for matrix $X$, which is accessed one row at a time. However, as the row size of $X$ increases and the matrix becomes less sparse, the locality of access for $X$ deteriorates due to the random access patterns dictated by the sparse row indices.

Furthermore, the separation of positive and negative value processing creates two distinct inner loops, which can disrupt cache performance when $K$ is large. Importantly, since the sparse matrix $W$ contains only ternary values $\{-1, 0, +1\}$, the algorithm naturally performs only additions and subtractions - no multiplications are required. Additionally, a bias vector $b$ of size $N$ is added to each row of $Y$.

\mypar{Implementation Note} Our scalar optimizations focus on the core $Y = XW + b$ operation, excluding PReLU activation to:
\begin{itemize}
    \item Maintain clear optimization targets (memory access patterns)
    \item Avoid unnecessary complexity in variant analysis
\end{itemize}
PReLU is included in vectorized implementations where activation fusion provides practical benefits for inference.

\mypar{Cost Analysis}
The cost can be measured based on the number of floating-point operations. Since the multiplication by \(\pm 1\) is implemented as additions and subtractions, our cost metric is the total count of floating-point additions (\texttt{fadds}).
Based on this instruction mix, the upper bound for performance on Apple M1 is 4 flops/cycle.

Therefore, the total computational cost \(C\) of the baseline algorithm is:
\[
C(M, K, N, s) = M  N  \left(1 + sK\right)
\]
where $s\in\{\frac{1}{2}, \frac{1}{4}, \frac{1}{8}, \frac{1}{16} \}$ is the fraction of non-zero elements in W, which we refer to as sparsity from now on.

\section{Optimizations Performed}\label{sec:yourmethod}


This section presents our systematic approach to the optimization process. We begin with loop unrolling to increase ILP, then address memory locality through blocking, and finally implement interleaving to optimize memory access patterns. Though some optimizations were explored concurrently during the initial phase of the project, they will be presented as they were combined sequentially. Throughout the project, empirical performance analyses guided parameter selection.

\mypar{Compiler}
Our first optimization was to determine the compiler and compiler flags leading to the highest speedup. We researched different combinations of flags and compilers that could contribute to a speedup, and then compared results from several different combinations. Prior to optimizations, we found that \texttt{g++ -O2 -march=native -fstrict-aliasing -DNDEBUG} was the best command, leading to an almost 2x speedup over a base \texttt{g++} command. Retrying the command after all optimizations were performed, we found a slightly higher speedup with \texttt{-O3 -march=native -fstrict-aliasing -DNDEBUG}.

\mypar{Loop unrolling}
Since the loop length depends on the input data, the compiler can't unroll the loops automatically. Therefore, we unroll to reduce loop logic instructions and increase ILP.

In this step, we also targeted the reduction of Write-After-Write (WAW) dependencies by employing multiple accumulators to increase ILP. 

We began by unrolling the innermost loop, where the baseline implementation used a single accumulator (\texttt{y\_val}). Our templated \texttt{UnrolledTCSC} implementation takes an \texttt{UNROLL\_FACTOR} parameter to control the degree of unrolling. Through systematic profiling, we determined that an unrolling factor of 12 consistently provided optimal performance across various input configurations.

During our exploration of different sparse formats, we discovered that unrolling the outer loop in addition to the inner loop further improved performance. While this approach increases the working set size (as multiple rows of matrices X and Y must be kept in memory simultaneously), the performance benefits outweigh the locality costs.

\begin{figure}[htbp]
	\centering
	\includegraphics[width=0.5\textwidth]{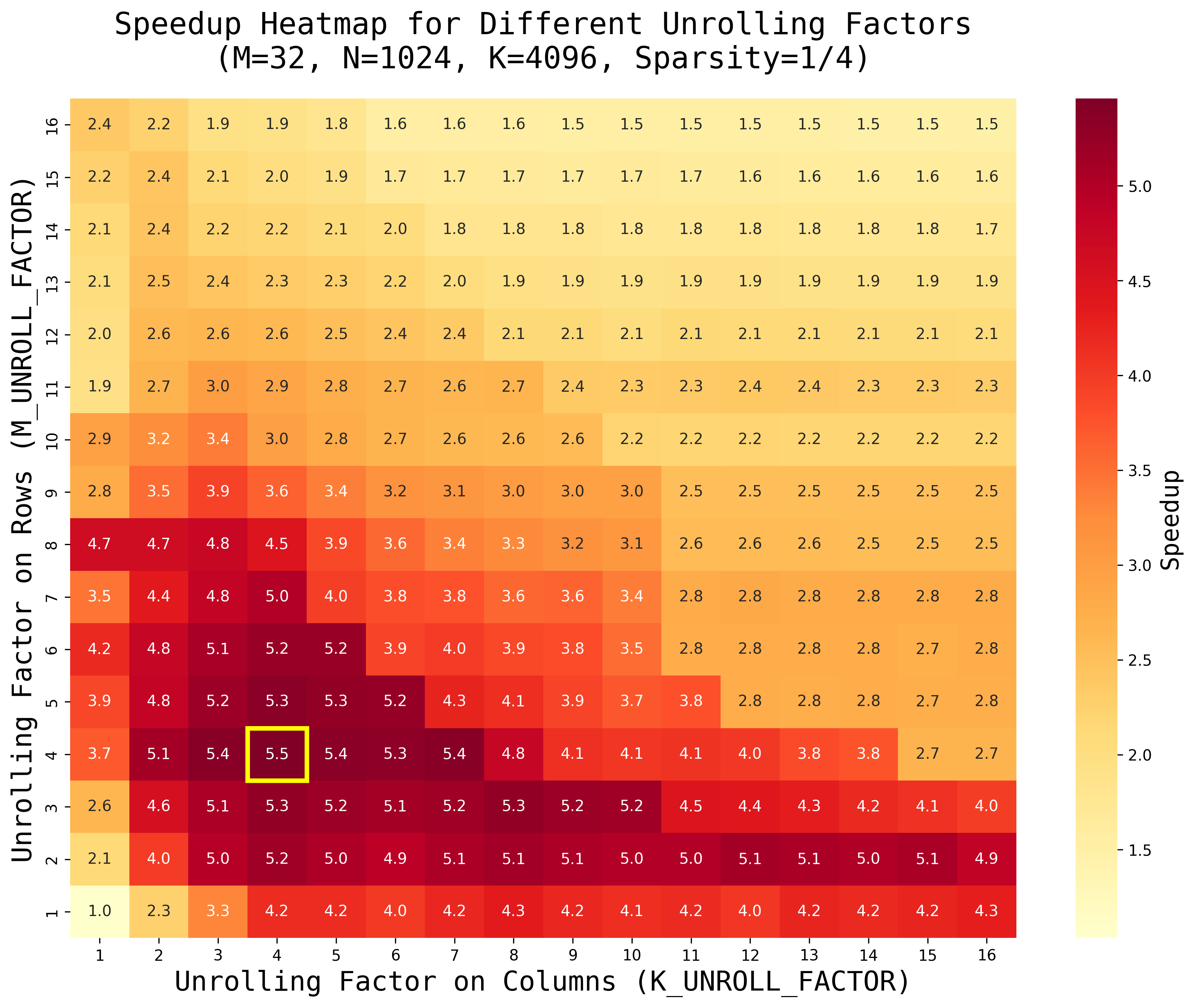}
    \caption{For any $K\leq4096$, we obtain the same optimal unrolling factor.}
        \label{fig:hm1}
\end{figure}

\begin{figure}[htbp]
	\centering
	\includegraphics[width=0.5\textwidth]{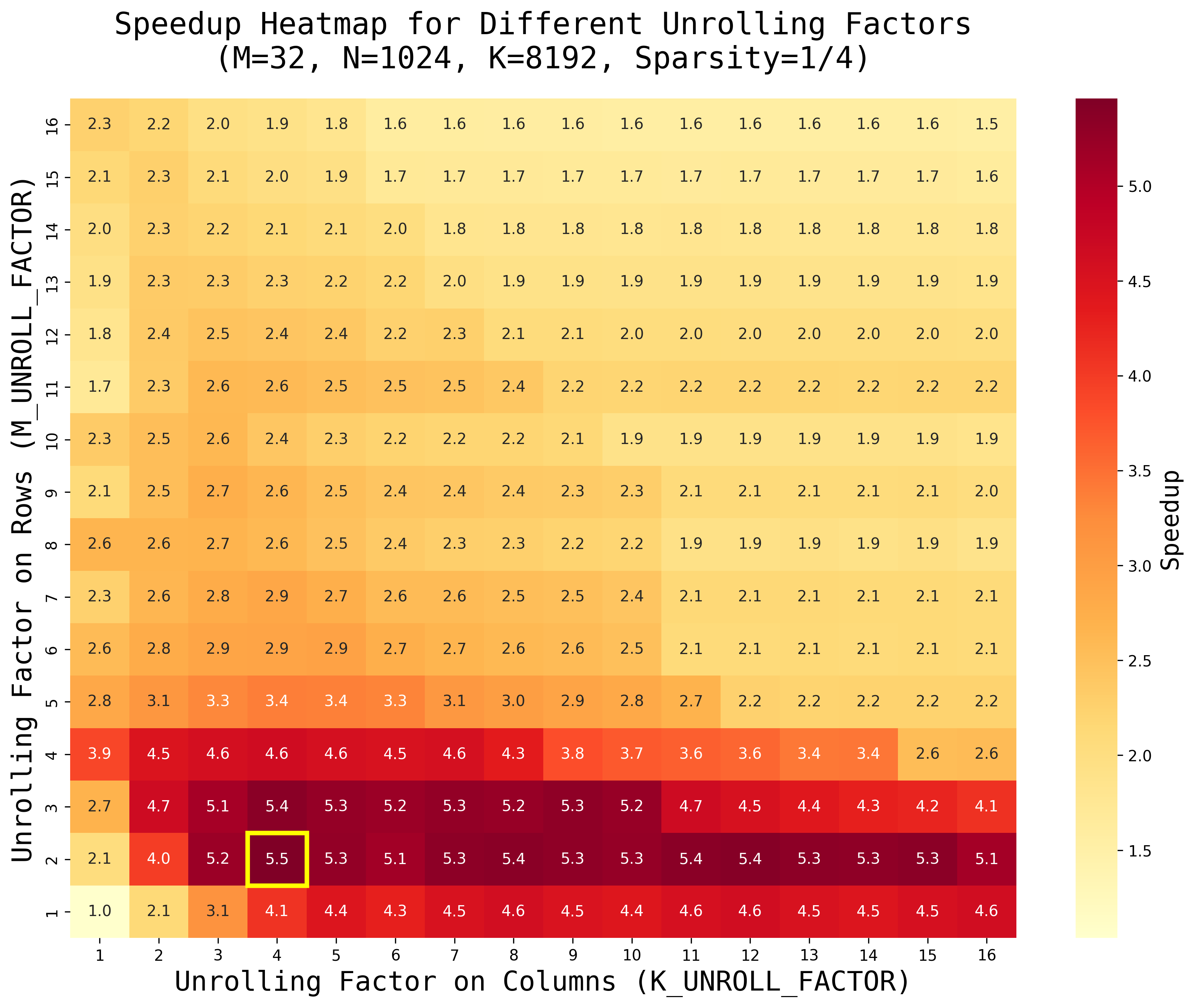}
    \label{fig:speedup8192}
    \caption{As the size of the row increases, cache misses occur when the working set is 4 rows of Y and X.}
\label{fig:hm2}
\end{figure}

\begin{figure}[htbp]
	\centering
	\includegraphics[width=0.5\textwidth]{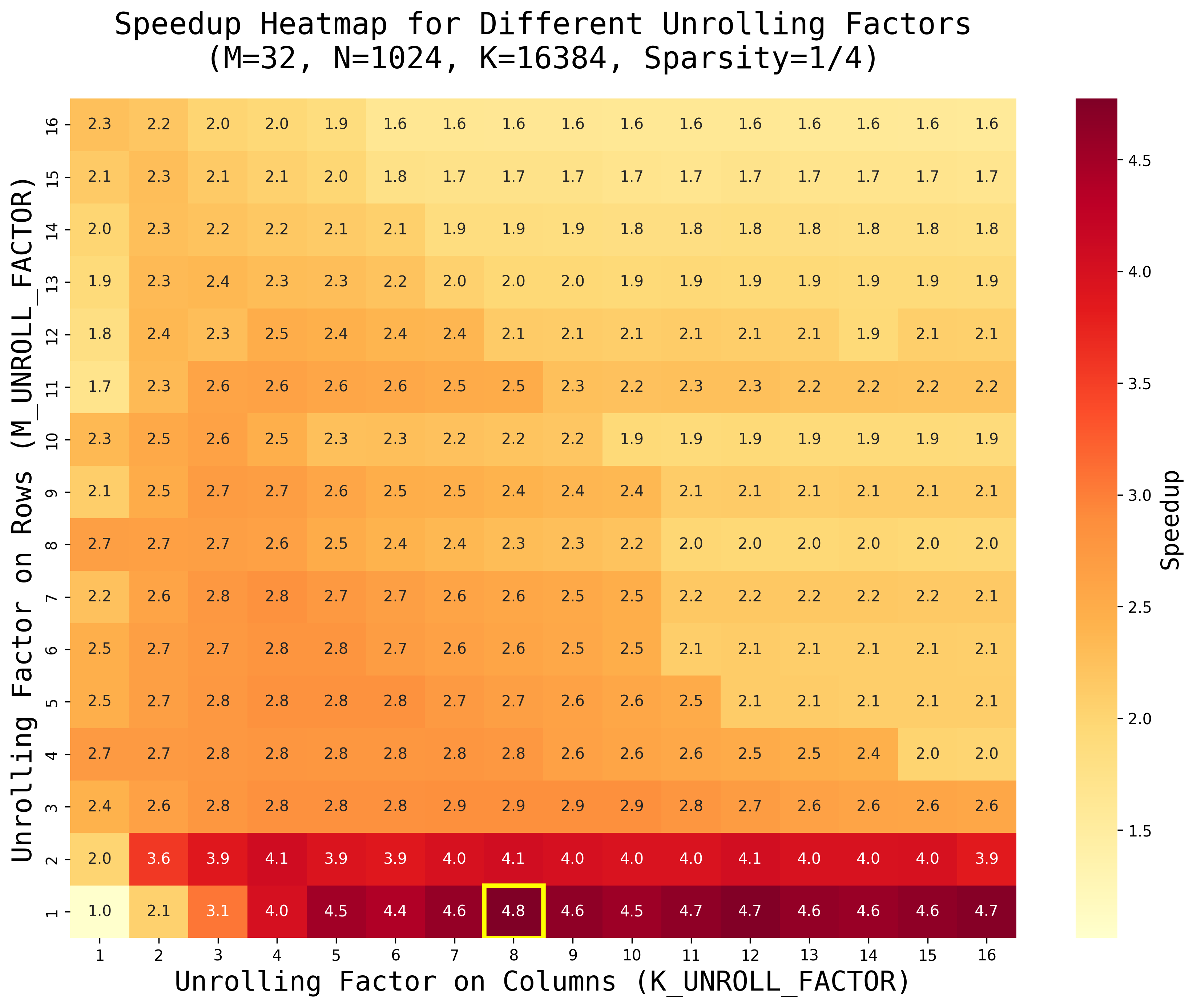}
    \label{fig:speedup16384}
    \caption{For the largest K, only one row of X and Y fits in cache.}
\label{fig:hm3}
\end{figure}

To systematically identify optimal parameters, we conducted a comprehensive grid search across different input sizes. We fixed the sparsity at 25\%, M=32, and N=1024, as preliminary benchmarks revealed these parameters had minimal impact on performance compared to K. We then varied K from 1024 to 16384 in powers of 2. The resulting speedups compared to the baseline implementation are shown in the heatmaps of Fig~\ref{fig:hm1}, Fig~\ref{fig:hm2} and Fig~\ref{fig:hm3}.


The results show that optimal unrolling factors shift toward lower values as K increases. This occurs because the working set of X consists of M rows, each containing K elements. For K $< 4096$, we can fit 4 complete rows in cache, while K=8192 and K=16384 begin to experience cache misses. This observation motivated our next optimization, which focuses on improving cache locality to maintain consistent performance for the \texttt{UnrolledTCSC\_K4\_M4} version across larger K values.

\mypar{Blocking}
During our optimization process, we identified a critical locality issue in the baseline TCSC format. While matrices X and Y exhibit good spatial locality with row-by-row access patterns, matrix X elements within a row are accessed according to row index vectors stored in the sparse format. This creates a practically random access pattern, since row indices are only sorted within individual columns of the sparse matrix W, but can span the entire range [0, K) within each column.

In the baseline TCSC format, for each column j, we first process all positive row indices, then all negative row indices. This means that when accessing X[row\_index], we might jump from X[0] to X[K-1] and back, leading to poor spatial locality and cache misses when K is large.

To address this locality issue, we developed a blocked sparse format (\texttt{BlockedTCSC}) that fundamentally reorganizes how row indices are stored and accessed. Instead of processing all K rows for each column sequentially, we divide the K rows into blocks of size B and process the data block-by-block.

We change the iteration order from:
\begin{itemize}
    \item \textbf{Baseline}: For each column j, process rows [0, K)
    \item \textbf{Blocked}: For each block b, for each column j, process rows [b×B, (b+1)×B)
\end{itemize}

This reorganization ensures that row indices within each processing phase are constrained to a range of size B, dramatically reducing the working set size for matrix X from K elements to just B elements at any given time.

The blocked format stores data as follows: for each block b (where b = 0, 1, ..., K/B-1), we store column start positions and row indices for all columns within that block range. This means that when processing block b, all accessed elements X[row\_index] have row\_index in the range [b×B, (b+1)×B), ensuring spatial locality.
\begin{figure}[htbp]
	\centering
	\includegraphics[width=0.5\textwidth]{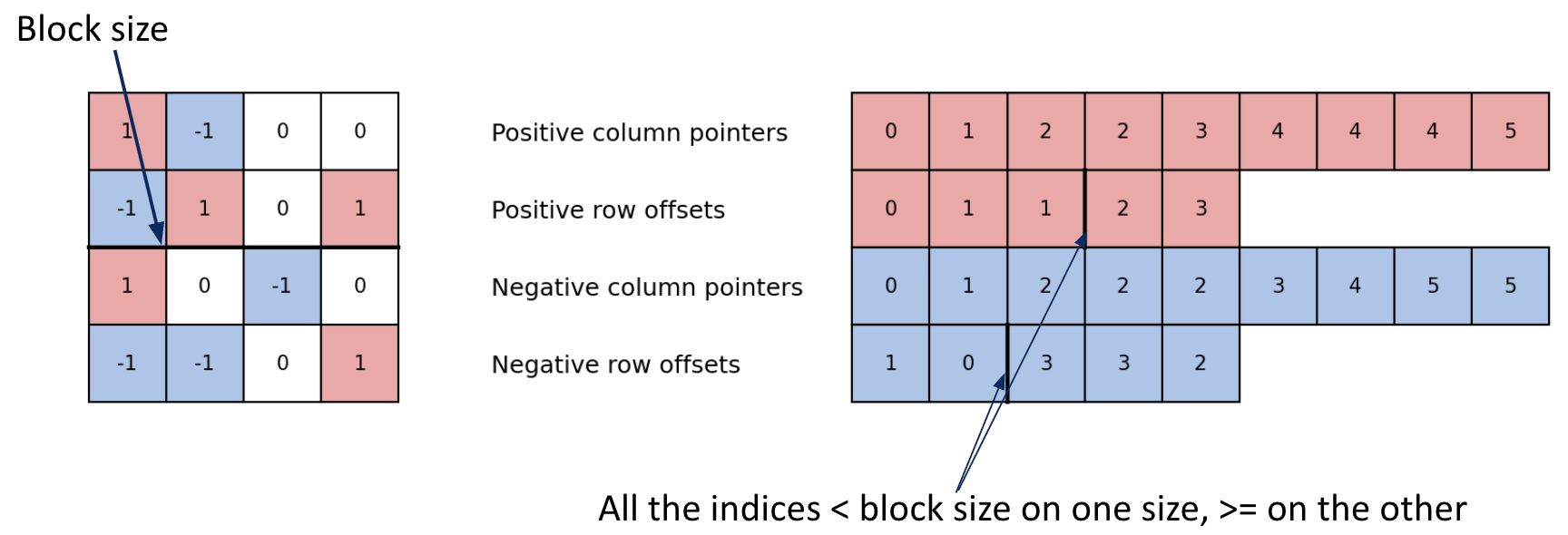}
	\caption{Example of BlockedTCSC format for B=2}
	\label{fig:blocked_tcsc}
\end{figure}

As illustrated in Fig~\ref{fig:blocked_tcsc}, all elements of X[i] accessed during the first phase belong to the range [0,B), while elements accessed during the second phase belong to [B,2B). Although this approach disrupts the sequential access pattern for Y (requiring each Y element to be accessed multiple times across different blocks), the significant improvement in X's spatial locality more than compensates for this trade-off, especially for large K values.

We applied this blocked format to address the locality challenges for larger K values in our \texttt{UnrolledTCSC\_K4\_M4} implementation. Through empirical testing, we determined the optimal block size to be 4096 elements. This choice is well-motivated, as 4096 represents the largest K value for which we could fit 4 complete rows in cache without experiencing significant cache misses, as we saw in the previous section.

\begin{figure}[htbp]
	\centering
	\includegraphics[width=0.5\textwidth]{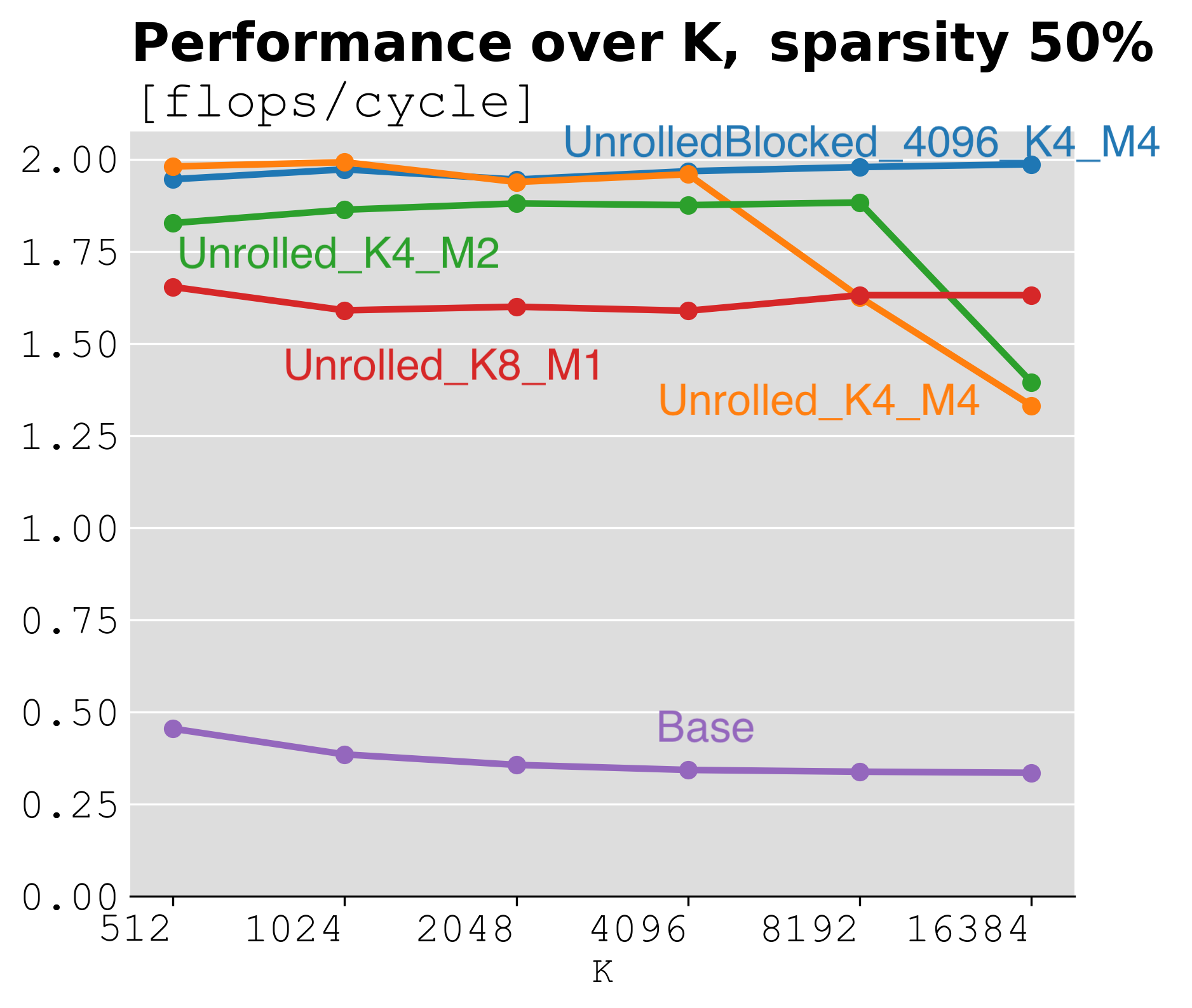}
	\caption{Performance in flops/cycle across different \( K \) values for various kernel variants at 50\% sparsity. Each line represents a different function.}
    \label{fig:perf_best_unrolling}
\end{figure}

The performance comparison between our \\
\texttt{UnrolledBlockedTCSC\_K4\_M4} implementation and the fastest unblocked versions for each K value is presented in Fig~\ref{fig:perf_best_unrolling},

The results show that the
\texttt{UnrolledBlockedTCSC} version maintains consistent performance across all K values, preserving the same level of performance achieved for smaller input sizes.

\mypar{Interleaving}
The baseline algorithm stores the locations of positive and negative ones in separate data structures (\texttt{row\_index\_pos} and \texttt{row\_index\_neg}). This separation creates irregular memory access patterns for matrix X: although accesses within each individual inner loop are sorted, the algorithm must execute two separate inner loops, disrupting spatial locality on X and potentially causing cache misses between processing positive and negative indices.

We explored several approaches to address this issue, including storing signs as sign bits within the row index vector (inverted indices format). Our successful solution implements an interleaved storage format that fundamentally changes how row indices are organized.

We first developed a basic \texttt{InterleavedTCSC} format that addresses the separate loop issue. For each column, we collect positive and negative indices separately, then interleave them in variable group sizes. After fine-tuning the group parameter based on testing, we used groups of 4. This meant the values array alternated between four positive indices and four negative indices. Remaining unmatched indices for each column are stored separately. This eliminates the need for separate positive/negative loops while maintaining spatial locality within each column.

\begin{figure}[htbp]
	\centering
	\includegraphics[width=0.5\textwidth]{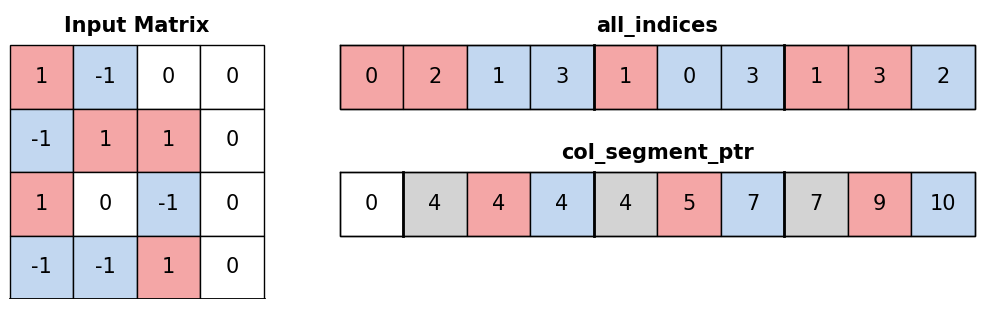}
	\caption{Example of InterleavedTCSC with grouping size = 2}
	\label{fig:interleaved_tcsc}
\end{figure}

\mypar{Interleaving + Blocking}
Building on this foundation, our \texttt{InterleavedBlockedTCSC} format combines blocking with interleaved storage. Instead of separate positive and negative arrays, we use:
\begin{itemize}
    \item \texttt{all\_indices}: Single vector storing all row indices in interleaved groups
    \item \texttt{col\_segment\_ptr}: Pointers marking three segments per blocked column: interleaved pairs, remaining positives, remaining negatives
\end{itemize}

For each column within a block, we collect positive and negative indices separately, then interleave them in groups. With unroll factor $F$, we compute $F/2$ positive indices followed by $F/2$ negative indices. Any remaining unmatched indices are stored separately at the end.

The algorithm processes each column in three phases: first the interleaved pairs, then remaining positives, then remaining negatives. For our implementation, we empirically concluded that the optimal grouping size was 4 indices for both +1s and -1s.
Our solution, for big sizes and high densities that allow the additional branching due to the management of cleanup code to be worthwhile, performed consistently better than both the vanilla and the blocked formats.

\mypar{SIMD Vectorization}
In our SIMD vectorization efforts, we create ``symmetry" in our algorithms, and then exploit this using SIMD instructions that perform more flops per cycle (four times more, in our case).

First, we vectorized the baseline algorithm in two ways. Both ways involve computing four row-elements of Y per iteration and use a ``symmetric" version of the interleaved data structure for pairs of signs.
Symmetry is achieved by mandating that every four columns of W store the same number of interleaved indices, that it is a multiple of 4, and that if a sign is in excess, then the pair indices indicating a deficit sign actually point to a dummy value of O, thus having no effect in the sum calculation.

In our ``vertical" approach, we use one slot of the vector register per row-element of Y. The high-level goal is to multiply each row of X with 4 columns of W in every iteration, thus K must be divisible by 4. In each innermost iteration, we process 2 groups of signs for each column (so 4 values), aggregating the partial sums in one positive and one negative sum vector register.
In the end, we subtract these registers and we obtain the final values of Y in each slot of the outcome vector register.

In our ``horizontal" approach, the high-level goal is again to multiply each row of X with 4 columns of W in every iteration. The difference with the former approach is that we use 1 vector register per column in which we keep the partial sums, and in the end we horizontally add the slot values to obtain the final Y value.

Both of the aforementioned approaches are expected to perform similarly, since they have a similar flop count with the baseline, excluding some padding needed to create symmetry, and they perform 4 flops per vector instruction.

Our last vectorization approach aimed at improving our best scalar version, which is an implementation blocked in 4096 values, interleaved in groups of two values per sign, unrolled on four rows of M and four columns of K. We vectorized on M and K, processing four values of K (columns of W) for each of four rows of M (rows of X) per iteration.
We use four vector registers, one for each column of W, with each of the four slots mapping to one row of X. We left the cleanup code for the values that could not be grouped and the excessive signs to its scalar implementation. 

We observe that our vectorized implementation of our optimal scalar implementation performs similarly, but not better. This is because sparse GEMM makes heavy use of non-sequential memory reads. The NEON instruction set does not provide a ``gather" vector instruction for random memory operands. ARM scatter/gather instructions are provided in ARM SVE, which is not supported on Apple Silicon. Though we experimented with serializing memory accesses by first copying to a local buffer, the overhead completely hindered the performance gains of sequential vector loads.

\mypar{Value Compression}
An optimization for compressing the \(W\) matrix—packing five ternary entries into one byte—was prototyped but omitted from the final kernel. In TCSC each ternary value is stored as a 32-bit index, so to encode \(t\) adjacent values in \(b\) bits we require $2^b \geq 3^t$, and we minimize $\frac{2^b - 3^t}{2^b}$. Focusing on small \(t\) (to avoid wasted computation on grouped zero values) and ``convenient" \(b\) (fitting in common data primitives), we found that \(t=5\) and \(b=8\), gave only 5.08 \% wasted space and representing each 5-tuple in one \texttt{uint8\_t}. At load time we treat each group of five \(\{-1,0,+1\}\) values as a 5-digit base-3 number and convert it to an 8-bit code. In the compute loop we decode via a 243-entry lookup table (\(\texttt{uint8\_t}\to\) five‐tuple) that fits in L1 and costs zero flops.

When compared to the baseline structure unrolled by 5, this approach gave a speedup at sparsity 50\%, matched baseline at sparsity = 25\%, and underperformed for lower sparsity values due to wasted work on zero values. Splitting into groups of 2 and 5 reduced some wasted work but introduced runtime branching or harmed spatial locality. The overhead introduced by the extra complexity outweighed the benefits, so value compression was dropped from the final optimized kernel.


\mypar{Inverted Index}
Baseline \texttt{TCSC} stores positive and negative row indices in separate arrays, forcing two passes over~$X$ and hurting cache locality. The inverted‐index approach merges them into a single vector by encoding $+1$ as index~$i$ and $-1$ as \(\sim i\) (bitwise NOT), halving the number of column pointers and unifying the inner loops. However, the cost of branching when decoding both the index and its sign in the innermost loop proved too high, reducing the overall flops‐per‐cycle count compared to the baseline. We therefore abandoned it in favor of more effective strategies like blocking and interleaving.

\section{Experimental Results}\label{sec:exp}

\mypar{Experimental setup} The objective was to optimize for Apple M1, M2, and M3 processors. The final benchmarks were run on the M1, which features four high-performance cores at up to 3.2 GHz with a 128 KB L1 data cache and a 12 MB shared L2 cache. Apple Clang version 16.0.0 was used with the flags \texttt{-O3 -march=native -fstrict-aliasing -DNDEBUG}. The peak attainable performance of the processors for scalar code is 4 flops per cycle, whereas for vectorized code it is 16 flops per cycle. As a floating point operation we count both additions and multiplications.

\begin{figure}[htbp]
	\centering
	\includegraphics[width=0.5\textwidth]{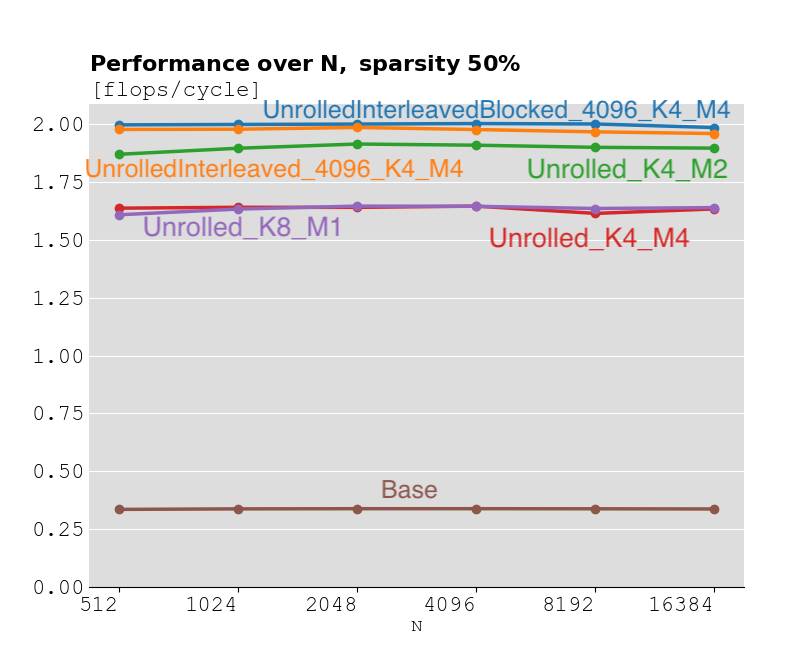}
    	\caption{Fixing K=8192, M=8 performance remains constant across different N values.}
	\label{fig:varying_N}
\end{figure}

\mypar{Results}
Our complete benchmark suite encompasses test cases with varying values of M, K, N, and sparsity. To present a meaningful and tractable performance analysis, we first identified which parameters significantly impact performance.
We determined that both M and N have negligible effects on performance. This behavior is expected: M represents the number of output rows and corresponds to external loop iterations that do not alter the algorithm's working set or memory access patterns. As shown in Fig~\ref{fig:varying_N}, performance remains consistent also across different N values.

Based on this analysis, subsequent results focus on the critical parameters K and sparsity, fixing M and N to convenient values in order to have an acceptable execution time for benchmarks.

\begin{figure}[htbp]
	\centering
	\includegraphics[width=0.4\textwidth]{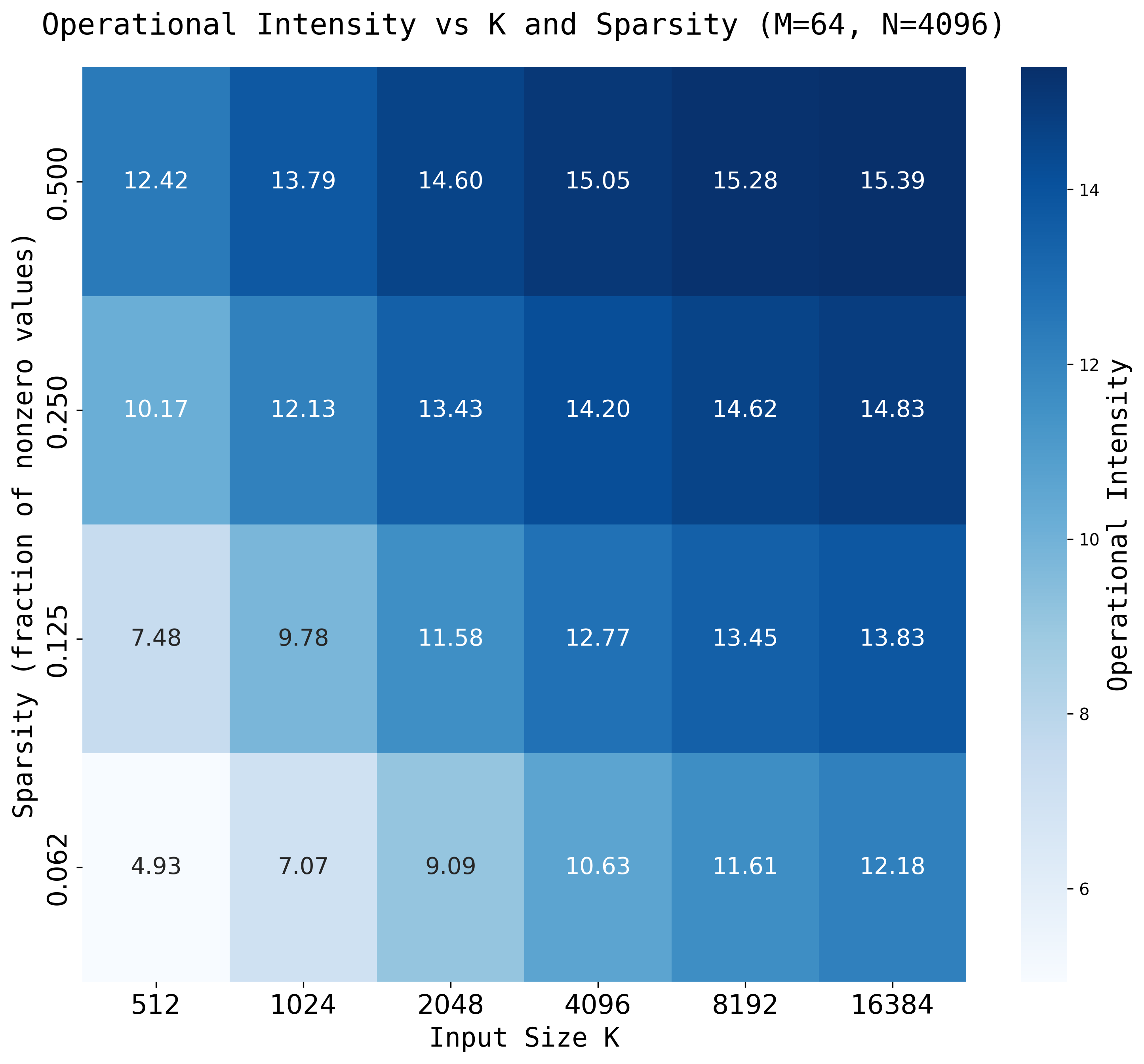}
	\caption{Operational Intensity heatmap for BaseTCSC for different sizes of K and different sparsity values. The estimate amount of input data is calculated based on the exact size of the sparse format, X, Y and the bias vector b.
    }
    \label{fig:oi}
\end{figure}

\mypar{Different sparsity values}
\begin{figure}[htbp]
	\centering
	\includegraphics[width=0.4\textwidth]{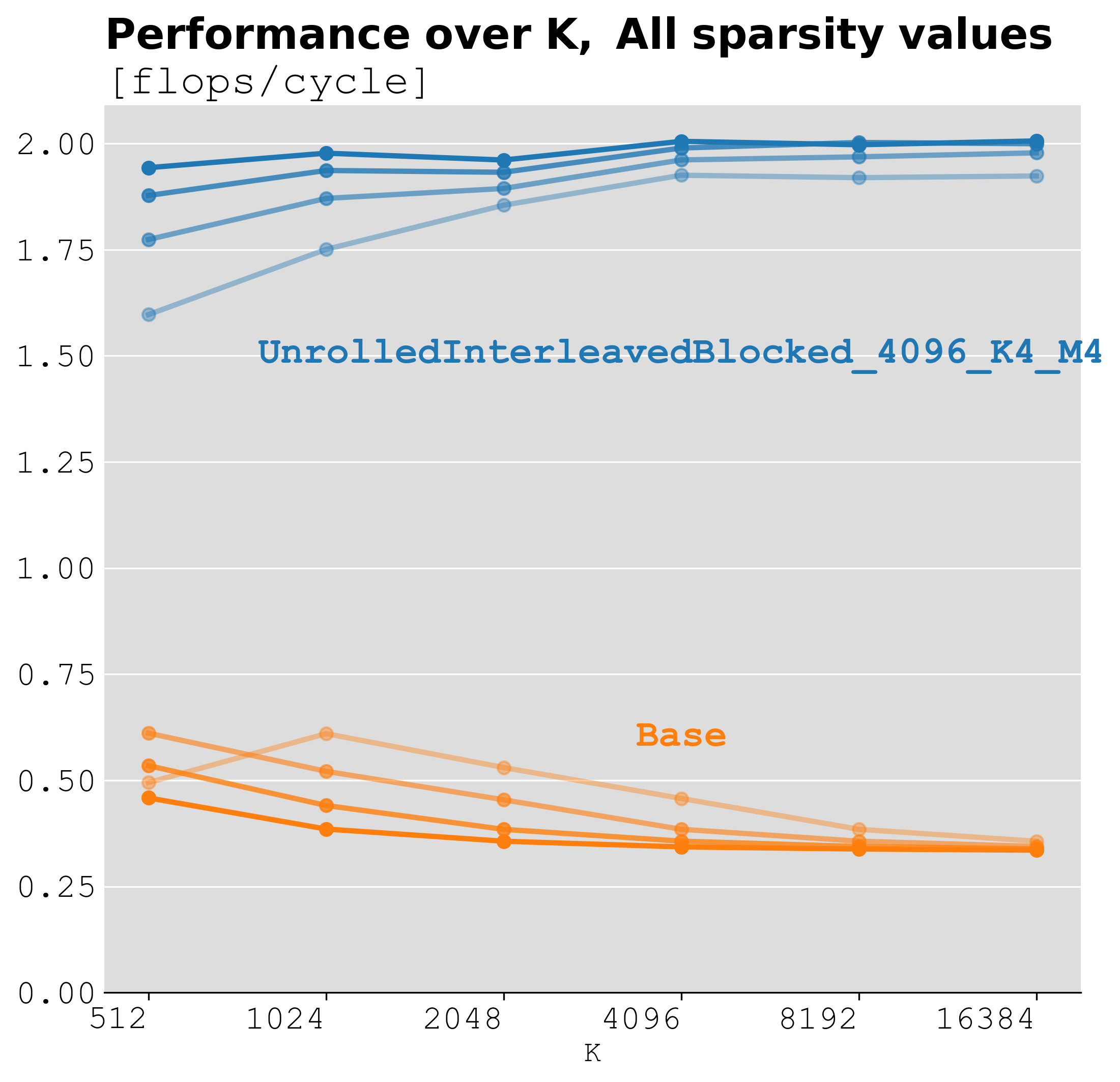}
	\caption{Performance over K for M=64, N=4096, for sparsity values of 50\%, 25\%, 12.5\% and 6.5\%. More transparent lines denote lower sparsity. Block Size is $\min(K, 4096)$.
    }
    \label{fig:all_sparsities}
\end{figure}
We measured the change in our accelerated function's performance as sparsity varied. In Fig~\ref{fig:all_sparsities} we compare the fastest implementation for sparsity values of 50\% (solid lines), 25\%, 12.5\% and 6.5\% (least opaque lines). Our best scalar implementation scales very well with respect to K, maintaining steady performance for $K \geq 4096$ for all sparsity levels. This suggests increasing K doesn't introduce new non-compulsory misses on X and Y. The baseline implementation achieves at best a 15.3\% percentage of peak performance for $K=1024, s = 6.5\%$. Our implementation achieves 50.2\% of peak performance for $K=16384, s=50\%$. For $K=16384, s=50\%$ we observe the greatest increase in performance of our implementation compared to the baseline, with a 5.98x increase.
Moreover, it is evident that the performance of the most optimized version for lower sparsity values and smaller values of K is lower compared to higher sparsity values and larger Ks.
In Fig~\ref{fig:oi} we show how the operational intensity follows the same trend.
The fact that in all the cases where the operational intensity is lower the performance also happens to be lower makes us confident that the computation is memory bound and that the performance obtained could not be improved more. On the other hand, the absence of an insightful cache analysis tool for Apple Silicon processors makes it difficult to obtain a better estimate of the actual memory movement.

\begin{figure}[htbp]
        \centering
	\includegraphics[width=0.5\textwidth]{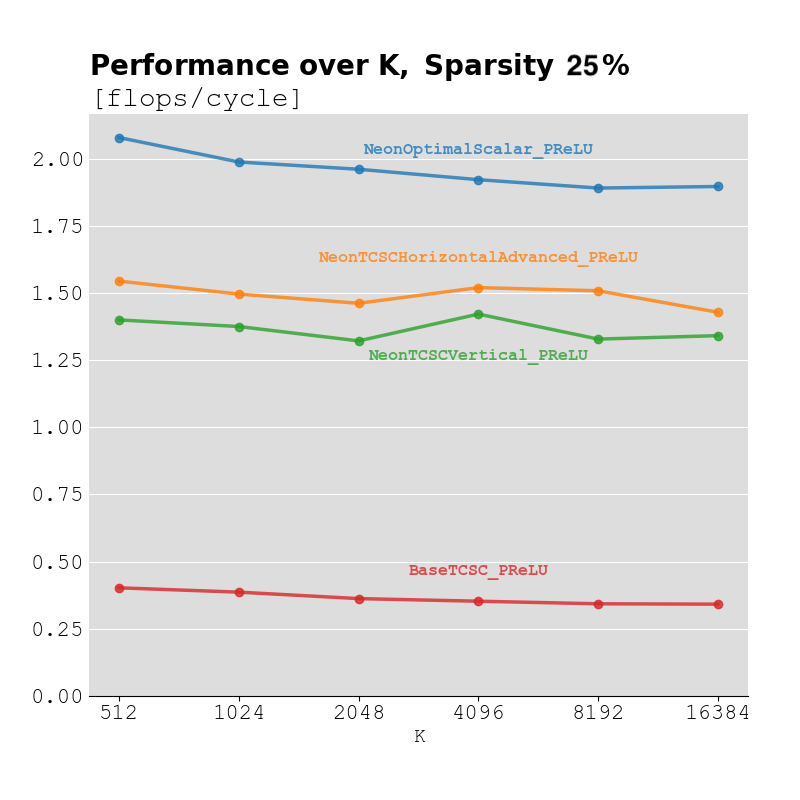}
	\caption{Performance over K of vectorized implementations for fixed M=N=1024 and 25\% sparsity. Horizontal and Vertical achieve about 3.5x performance increase, the optimal scalar implementation achieves about 5x performance increase.}
        \label{fig:vector}
\end{figure}

\mypar{Vectorization results}
Performance of our vectorized implementations are shown in Fig~\ref{fig:vector}, with BaseTCSC as the baseline comparison. The PReLU computation is implemented in all of the plotted vectorized functions. The optimal scalar implementation is blocked and interleaved. Its block size is defined as $\min(K, 4096)$. We only visualize results for $s=25\%$, as results were consistent across other sparsity values. The results illustrate that our horizontal and vertical implementations perform similarly, with a 3.5x in performance increase over the baseline. This is close to the theoretical peak 4x increase. The optimal scalar implementation achieves over a 5x performance increase, likely because ILP is utilized in the scalar part of the cleanup code.
We observe that all implementations are scalable, since there are no fluctuations in performance despite the increase of K. For $K=512$ we observe the greatest increase in performance of our vectorized implementations compared to the baseline, at 5.59x. 

\section{Conclusions}



Sparse Ternary GEMM on Apple M1 demands both high ILP and tight memory locality to consistently achieve peak performance. We found that the main speedup was attained by unrolling both the inner and the outer loop. To fix the increase of the working set, blocking the data format was needed. The last minor speedup was obtained by switching to an interleaved sparse format. Our kernel sustains a 5.98× speedup at 50 \% sparsity (50.2 \% of attainable peak processor performance). Notably, the scalar implementation we present achieves better performance than the vectorized implementation on Apple Silicon, due to the absence of scatter/gather instructions.


\end{document}